\documentclass[twocolumn,showpacs,preprintnumbers,amsmath,amssymb]{revtex4}

\usepackage{graphicx}
\usepackage{dcolumn}
\usepackage{bm}

\begin{document}

\preprint{}

\title{Erasing Distinguishability Using Quantum Frequency Up-Conversion}

\author{Hiroki Takesue$^{1,2}$} \email{htakesue@will.brl.ntt.co.jp}
\affiliation{%
$^1$NTT Basic Research Laboratories, NTT Corporation, 3-1 Morinosato Wakamiya, Atsugi, Kanagawa, 243-0198, Japan\\
$^2$CREST, Japan Science and Technology Agency, 4-1-8 Honcho, Kawaguchi, Saitama, 332-0012, Japan
}%

\date{\today}

\begin{abstract}
The frequency distinguishability of two single photons was successfully erased using single photon frequency up-conversion. A frequency non-degenerate photon pair generated via spontaneous four-wave mixing in a dispersion shifted fiber was used to emulate two telecom-band single photons that were in the same temporal mode but in different frequency modes. The frequencies of these photons were converted to the same frequency by using the sum frequency generation process in periodically poled lithium niobate waveguides, while maintaining their temporal indistinguishability. As a result, the two converted photons exhibited a non-classical dip in a Hong-Ou-Mandel quantum interference experiment. The present scheme will add flexibility to networking quantum information systems that use photons with various wavelengths.  

\end{abstract}

\maketitle

In recent years, remarkable progress has been made on photonic quantum information systems \cite{kok}. Several important quantum information experiments, such as controlled-NOT gates \cite{pittman,cnot} and one-way quantum computation \cite{walter}, have already been reported. Most of those experiments were undertaken using photons in a short wavelength band (around 800 nm), because good single photon detectors are available in these bands. 
On the other hand, schemes for generating entangled photon pairs in the 1.5-$\mu$m telecom band have been intensively studied over the last several years \cite{yoshizawa,takesue,li,takesueppln,silicon}. The entangled photons generated with those schemes are suitable for long-distance distribution over optical fiber networks. A long-distance entanglement distribution over 100 km of optical fiber has already been demonstrated \cite{honjo,qiang}. Distributed entangled photon pairs over optical fiber networks can be an important resource for networking quantum information systems. For example, we can integrate several photonic quantum information systems installed at distant locations into a larger quantum system using quantum teleportation \cite{qt}. 
However, 1.5-$\mu$m entangled pairs cannot be directly employed for teleporting short wavelength photons, because a Bell state measurement (BSM) using a beamsplitter, which is an essential component of the quantum teleportation experiment reported in \cite{qt}, does not work on two frequency-distinguishable photons. 

We can possibly overcome this problem by using quantum frequency up-conversion based on the sum-frequency generation (SFG) process \cite{kumar}. After the first experimental demonstration \cite{huang}, this technology has been used as a way to convert a 1.5-$\mu$m telecom-band photon that is suitable for fiber transmission to a short wavelength photon that can be detected using a highly-efficient silicon avalanche photodiode (APD). These ``up-conversion detectors" for the 1.5-$\mu$m band were realized using bulk periodically poled lithium niobate (PPLN) crystals \cite{albota,van} and a PPLN waveguide \cite{carsten}, and were employed in several application experiments including high-speed and long-distance quantum key distribution \cite{njp,thew}, fast optical time domain reflectometry \cite{eleni,otdr2}, and entanglement measurements \cite{honjo,tanzilli}. However, the application of quantum frequency up-conversion has mostly been limited to up-conversion detector experiments.

In this paper, I describe a novel application of quantum frequency up-conversion: tuning the frequency of a pulsed single photon while maintaining its Fourier transform limited characteristics. 
The elimination of the frequency distinguishability of two single photons is successfully demonstrated using up-converters based on PPLN waveguides. 
As a result, the two up-converted photons exhibited a non-classical dip in a Hong-Ou-Mandel quantum interference experiment \cite{hom}. 
The presented scheme is expected to be an important basic technology that adds flexibility to networking quantum information systems using photons with various wavelengths.

First, let us briefly review quantum frequency up-conversion based on SFG. When the pump is strong and undepleted, the interaction Hamiltonian of the SFG process is given by \cite{kumar}
\begin{equation}
\hat{H}_I = i \hbar \chi (\hat{a}_L \hat{a}_S^\dagger - h.c.),
\end{equation}
where $\hat{a}_L$ and $\hat{a}_S$ are the annihilation operators for long and short wavelength photons, respectively, and $\chi$ is a constant that is proportional to the second-order susceptibility $\chi^{(2)}$. 
The Heisenberg equations of motion in the interaction picture are
\begin{eqnarray}
\frac{d \hat{a}_L}{dt} &=& - \chi \hat{a}_S \\
 \frac{d \hat{a}_S}{dt} &=& \chi \hat{a}_L.
\end{eqnarray}
The solutions to these equations are given by
\begin{eqnarray}
\hat{a}_L (t) &=& \hat{a}_{L0} \cos \chi t - \hat{a}_{S0} \sin \chi t, \\
\hat{a}_S (t) &=& \hat{a}_{S0} \cos \chi t + \hat{a}_{L0} \sin \chi t, 
\end{eqnarray}
where $\hat{a}_{x0}$ ($x=L,S$) are the annihilation operators for mode $x$ at $t=0$. 
The average number of SFG photons $\langle \hat{a}_S^\dagger \hat{a}_S\rangle = \langle \hat{n}_S \rangle$ is given by
\begin{equation}
\langle \hat{n}_S \rangle = \langle \hat{n}_{S0} \rangle \cos^2 \chi t + \langle \hat{n}_{L0}\rangle \sin^2 \chi t.
\end{equation}
For an up-converter, $\langle \hat{n}_{S0} \rangle=0$. 
This equation suggests that, unlike the difference frequency generation process, no spontaneous emission occurs in the SFG process. This means that the ideal frequency conversion of a single photon with no noise emission can be achieved with the SFG process.

\begin{figure}[thb]

\centerline{\includegraphics[width=\linewidth]{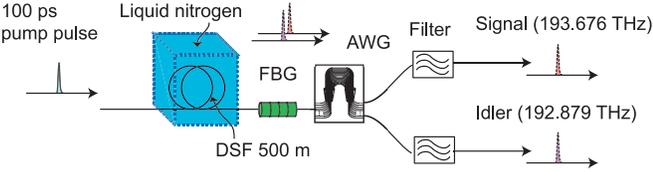}}

\caption{Correlated photon-pair source based on SFWM in DSF.}
\label{1}

\end{figure}

Figure \ref{1} shows the correlated photon-pair source with non-degenerate frequencies with which I emulated two single photons with the same temporal mode but with different frequencies. A 100-ps pump pulse with a wavelength of 1551.1 nm was input into a dispersion shifted fiber (DSF) in which broadband signal/idler photon pairs were generated by the spontaneous four-wave mixing (SFWM) process \cite{fio,takesue,li}. The DSF was soaked in liquid nitrogen to suppress the noise photons generated by the spontaneous Raman scattering process (SpRS) \cite{cool}. The photons output from the DSF were passed through a fiber Bragg grating (FBG) to suppress the pump photons, and launched into an arrayed waveguide grating (AWG), which is a waveguide-type grating that separates the signal and idler photons \cite{takahashi}. The bandwidths of the signal and idler channels had a full width at half maximum of 0.2 nm (25 GHz). 
The two AWG outputs were followed by bandpass filters to further suppress the residual pump photons. 
The center frequencies of the signal and idler photons, which were measured with an optical spectrum analyzer with a resolution of 0.9 GHz, were 193.676 and 192.879 THz, respectively. 
According to \cite{om}, we can observe ``quantum beating" when two photons with different frequencies are used as input photons for a Hong-Ou-Mandel type quantum interference experiment. However, since the frequencies of two photons differ by 800 GHz in this experiment, the beating period corresponds to a 1.25 ps relative delay time, which was difficult to resolve stably in the current experimental setup.

\begin{figure}[thb]

\centerline{\includegraphics[width=\linewidth]{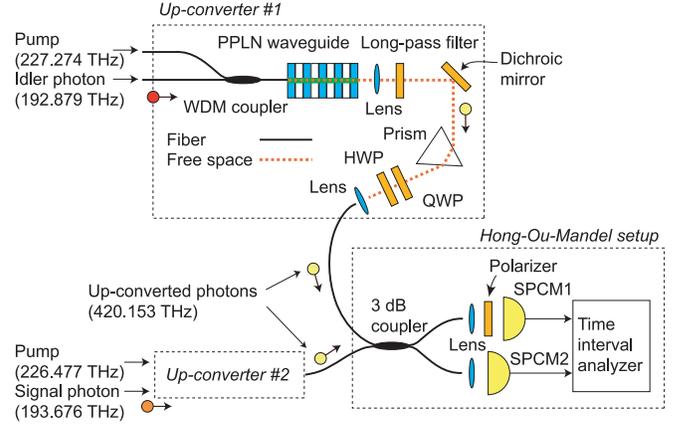}}

\caption{The up-converters and the Hong-Ou-Mandel quantum interference experiment setup.}
\label{2}

\end{figure}

\begin{figure}[thb]

\centerline{\includegraphics[width=\linewidth]{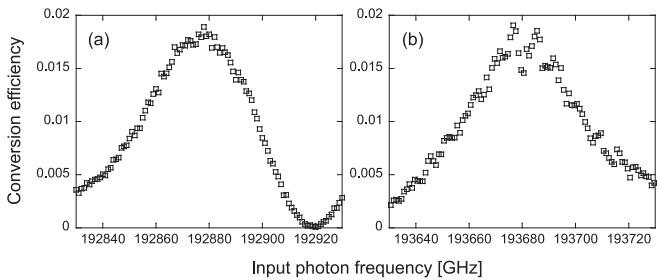}}

\caption{Conversion efficiency as a function of input photon frequency for the up-converters (a) for the idler and (b) for the signal. }
\label{3}

\end{figure}

Each photon from the source was passed through an optical delay line (Optoquest), and input into an up-converter, which is shown schematically in Fig. \ref{2}. The optical delay line was composed of fiber collimators and a corner cube reflector, and the delay was changed by adjusting the position of the corner cube reflector. 
In the up-converters, the input photons were combined using a strong 1.3-$\mu$m pump light with a wavelength-division-multiplexing (WDM) coupler, and launched into a PPLN waveguide, which was fabricated by HC Photonics. The SFG process in the waveguide up-converted the 1.5-$\mu$m photon to a 0.7-$\mu$m photon. The residual pump photons in the output from the waveguide were suppressed by a long-pass filter, a dichroic mirror, and a prism. Then, the photon was transmitted through a quarter wave plate (QWP) and a half wave plate (HWP), and collimated into an input port of a 3 dB coupler composed of single-mode fiber (SMF) for a Hong-Ou-Mandel quantum interference experiment \cite{hom}. 
The use of the SMF coupler enabled us to eliminate the spatial mode distinguishability between two up-converted photons. 
The two output ports of the coupler were connected to single photon counter modules (SPCM) based on Si APDs. As shown in Fig. \ref{2}, a polarizer was placed in front of SPCM1 to achieve the polarization indistinguishability. The HWPs and QWPs in the up-converters were adjusted so that the count rate of SPCM1 was maximized, and in this way the polarization distinguishability was eliminated. The frequencies of the pump lights for the signal and idler were set at 226.477 and 227.274 THz, respectively. With these pump frequencies, both up-converters had an SFG frequency of 420.153 THz. The detection signals from the SPCMs were used as the start and stop pulses for a time interval analyzer (TIA), and the coincidence between the signal and idler channels was counted. 
The pump powers input into the WDM couplers were set at 12.8 mW for the idler channel and 4.9 mW for the signal channel. At those pump powers, the combined dark count rate of the two SPCMs was approximately 4,000 cps. Most of the dark counts were caused by noise photons generated through spurious nonlinear effects such as SpRS \cite{carsten}. 
The conversion efficiencies of the up-converters as a function of photon frequency are shown in Fig. \ref{3}. Both up-converters had a peak conversion efficiency of around 2\% with a $\sim$40 GHz bandwidth. The fluctuation in the conversion efficiency for the signal channel (Fig. \ref{3} (b)) was probably caused by fluctuation in the domain inversion period of the PPLN waveguide.

\begin{figure}[thb]

\centerline{\includegraphics[width=.8\linewidth]{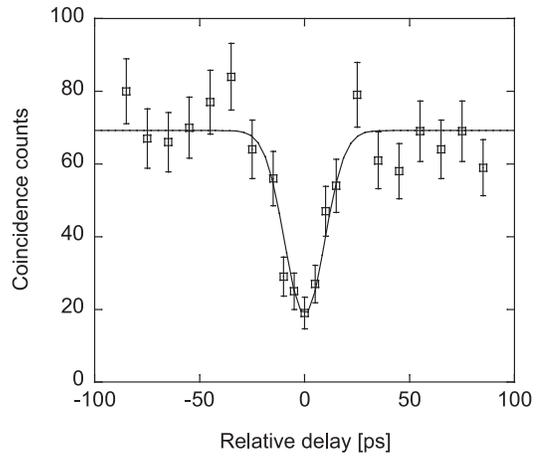}}

\caption{Experimental result. The horizontal axis shows the coincidence counts obtained by TIA for 500,000 start pulses. }
\label{4}

\end{figure}

I changed the relative delay between the signal and idler photons by changing the delay time provided by the optical delay line, and measured the coincidence counts for 500,000 start pulses. The average photon pair per pulse was set at approximately 0.05. The result is shown in Fig. \ref{4}. A clear dip was observed in the coincidence counts. The data shown in this figure are raw values without the subtraction of any accidental coincidences. 
When the photon pair has a Fourier transform limited pulse shape whose 1/e half width is given by $\sigma$, the shape of the dip can be fitted with the following function \cite{hom}. 
\begin{equation}
N_c = C\left\{ 1- V \exp \left(-\frac{\delta \tau^2}{2 \sigma^2} \right) \right\}
\end{equation}
Here, $V$, $\delta \tau$ and $C$ denote the visibility, the delay time, and a constant. 
The visibility of the fitted curve was 73.2$\pm$7.6\%, which clearly indicates that non-classical fourth-order interference was obtained. A 1/e half width of $9.5 \pm 1.3$ ps was obtained, which is in good agreement with the coherence time of a Fourier transform limited photon with a 25 GHz bandwidth, namely 10.6 ps. This result confirmed that the frequency distinguishability between two photons was successfully erased while maintaining their Fourier-transform limited characteristics using the single photon up-conversion process.

There are several possible reasons for the visibility being limited to 73\%. 
The first is the accidental coincidences caused by multi-pair emission from the photon pair source. The noise photons generated in SpRS in the DSF may also have caused the excess accidental coincidences, because a large portion of the output photons from the source was generated by SpRS even when the DSF was cooled at 77 K \cite{cool}. In addition, the relatively large dark count rates at the photon detection that were caused by the noise photons generated in the up-converters might have contributed to the excess accidental coincidences.

\begin{figure}[thb]

\centerline{\includegraphics[width=\linewidth]{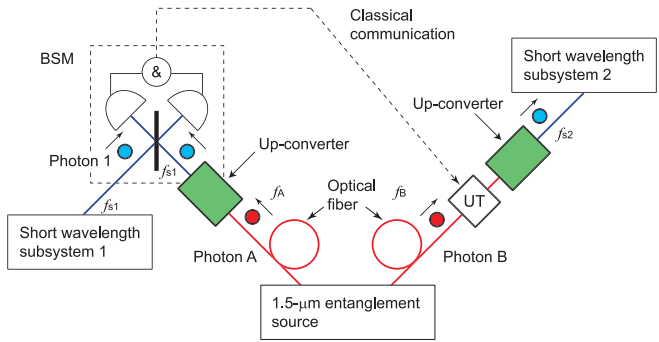}}

\caption{An example of quantum communication networking using quantum frequency up-conversion. UT: unitary transformation, BSM: Bell state measurement. }
\label{5}

\end{figure}

The obtained experimental result suggests the possibility of the flexible networking of quantum information systems with various wavelengths. 
An example is shown in Fig. \ref{5}. The system consists of quantum subsystems 1 and 2 that use short wavelength photons with frequencies of $f_{s1}$ and $f_{s2}$, respectively, and a 1.5-$\mu$m band entanglement source connected via optical fibers. 
Photons A (frequency $f_A$) and B ($f_B$) from the entanglement source are sent to subsystems 1 and 2, respectively. 
The frequency of photon A is up-converted to $f_{s1}$, and a BSM is undertaken on the up-converted photon and photon 1 output from subsystem 1 using a beamsplitter followed by single photon detectors. 
Then, by undertaking an appropriate unitary transformation to photon B using the BSM result sent by classical communication, the quantum state of photon 1 is transferred to that of photon B via the quantum teleportation protocol. By up-converting the frequency of photon B to $f_{s2}$, we can input the quantum state output from subsystem 1 to subsystem 2, and thus network two distant quantum subsystems. 
In this example, the temporal modes of the short-wavelength photon output from subsystem 1 and photon A in the 1.5-$\mu$m band should be indistinguishable. If we assume that the photons from the two subsystems and the entanglement source are generated via a spontaneous parametric process, temporal indistinguishability is achieved in the following way. 
The temporal shapes of the two photons can be matched by passing the photons through optical filters with the same spectral shape and the same frequency bandwidth for respective wavelength bands. 
Even when their temporal shapes are matched, temporal distinguishability (or timing jitter) is induced in the photon pairs if walk off occurs between the pump pulses and photon pairs in the spontaneous parametric process. Timing jitter can also occur if the pump pulse width is much greater than the coherence time of the photon pair. 
As in conventional quantum interference experiments using two independent photon-pair sources, we can overcome the first problem by using a short nonlinear medium to generate the entanglement, and can avoid the second by using pump pulses whose temporal width is shorter than the coherence time of the photon pairs \cite{rarity}. Thus, temporal indistinguishability between short and long wavelength photons can be achieved with existing technologies.

In summary, I have demonstrated the successful erasure of the frequency distinguishability of single photons by using quantum frequency up-conversion. A non-degenerate photon pair generated in a DSF was used as two single photons with frequency distinguishability. Then, the frequencies of the photons were converted to the same frequency by the up-converters based on SFG in PPLN waveguides, while maintaining the temporal indistinguishability. As a result, I observed a non-classical Hong-Ou-Mandel dip. The above scheme is expected to be useful for integrating photonic quantum information systems via quantum teleportation using 1.5-$\mu$m band entangled photons distributed over optical fiber networks.


\end{document}